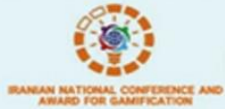

# Increasing the Diversity of Investment Portfolio with Integration of Gamified Components in the FinTech Applications' Lifecycle

Latifeh PourMohammadBagher*, Najmieh Sadat Safarabadi

Allameh Tabatab'i University, *Tehran*, Iran
*Department of Mathematics and Computer Science*

\* Corresponding author: Latifeh PourMohammadBagher, L_pmb@atu.ac.ir

**Abstract**

Gamification has the potential to make significant contributions to financial product delivery, Fintech services, and inclusive growth. The integration of gamification into FinTech applications has shown a positive correlation with the social impact theory. Utilizing gamification in a sustainable and effective manner can be crucial for long-term prospects in the FinTech industry. Therefore, it is essential to develop efficient and modern financial software that improves the customer experience. The current literature aims to contribute to this area by highlighting the relationship between interrelated theories and the key factors to consider when designing a gamified element. This study aims to explore the effects of gamification on altering user intention and its significant influence on customer value propositions.

## 1. Introduction

Fintech has been described as a "Financial sector innovation involving technology-enabled business models that can facilitate disintermediation [14]. The application of gamification in the FinTech domain delivers a value chain and "Gameful experiences support the users' overall value creation"[9],[1]. Introducing gamified elements and mechanisms into the Fintech apps brings new approaches to retaining customers. One of the main advantages of gamification is that it smoothens the process of learning and its curve while motivating the customer to take action and complete a particular task [1]. A gamified structure can have a positive influence on brand engagement and correct user engagement in a particular direction, which leads to solidifying customer loyalty and satisfaction [1].

It is important to distinguish between using games purely for entertainment and using game elements to gain real-life benefits. Gamification involves extracting and using the essence of games in non-gaming contexts to increase user awareness [10]. This relationship involves three main elements: internalization, compliance, and identification. Compliance refers to the customer value construct, which includes customer engagement, satisfaction, and loyalty. Identification refers to satisfaction and loyalty, while internalization relates to customer value outcomes [1]. In a gamified procedure, both customers and service providers alike can have considerable benefits [1]. Consequently, a FinTech platform can leverage all of these potentials. One of the main contributions of this literature is to propose a formulating strategy for gamification of the FinTech application with the goal of a more effective creation of customer value.

It is safe to say that currently, in the domain of Fintech software development, a gamified experience has not been a tangible procedure. The current paper provides contributions to the clarity of the facilitator's role in the gamified design choices and the role of research for each task-based procedure to achieve the desirable outcome in this field [6].

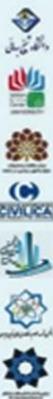
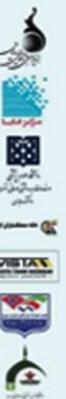

## 2. Previous Literature

The use of gamified elements in complex combinations or interactions has been limited so far. One previous application involved using gamification to improve user acceptance of new financial services and customer responses to innovative financial services. One example of this is the Robo-advisor, an automated investment solution that provides convenient portfolio balancing techniques, diversification techniques, and advanced customer experience. With such a system, traditional face-to-face financial advisory can be replaced with a more efficient approach [2].

To take a particular action, consumers form a perception of the available resources and support moreover the element of ownership motivates users to explore possible routes provided by FinTech software. While a fully AI-based and automated financial assistant is ambitious, elements from its horizon can be borrowed to modernize and automate certain tasks in the financial service. Facilitating conditions are important in developing game elements.

One of the aims is to learn about the effect of effort expectancy, ownership, and perceived value in combination and how they create a high degree of intention [3]. Visualizing user behavior through gamified functionalities assists and encourages them to stay in the specified network [5]. Virtual currencies or related assets form a strong ownership among users. According to this Functions such as receiving correspond to the importance of ownership in gamified elements.

The value associated with effort expectancy, facilitating conditions, and expected value have the highest level of effect in implementing a gamified element in FinTech platforms. These combinational motivational game-designed elements provoke behavioral intention to engage further with the FinTech platform and learn cooperatively [7]. Defining gamified elements based on combinational factors improves the user's life cycle within FinTech procedures.

According to the definition in alignment with gamification, it is a process of improving a service-based platform to reinforce the user's overall value creation [5]. It is important to note that the term gamification solely involves the game design facet or the experiential aspect that corresponds to the experience.

## 3. Fintech platform Life Cycle

The service life cycle of the Fintech applications includes basic service, information-based service, and advanced services. In such a platform the scheme of the modern advanced services strategy embodies assets management (AM), support for purchasing and selling stocks (ETFs), encouragement of digital cash transactions (NFTs), smart contract and peer-to-peer (P2P) lending, personal finance management by which users can see all of their finances in one place and including financial statement audits, consulting, technical accounting, tax consulting, and M&A activities [21], coverage for insureTech which leverages the use of technology to the insurance industry, and investment collection on business, foreign currency, industrial products. The aim of supporting many of these services in one place is to give individuals better financial decision-making, guidance, and capacity consequently assisting them to diversify their investment more effectively.

### 3-1- Influential factors in the gamification elements

Gamification involves developing game design elements for non-game contexts, incorporating components, mechanics, and dynamics. These game mechanics encourage specific behaviors and provide challenges and rewards based on performance, promoting interactions and elevating user engagement [1].

In the FinTech industry, gamification involves integrating game design features into financial and transactional processes, such as learning investment strategies, identifying opportunities, transferring and receiving payments, and tracking progress [2]. This can increase customer acceptance, satisfaction, and engagement, leading to customer loyalty and retention. Taking the Customer value factor into consideration is crucial in developing gamified elements, as it relates to perceived preference and evaluation of service attributes [9].

User engagement is also important, incorporating cognitive, behavioral, and emotional dimensions. Involving new marketing tools should consider effort expectancy, facilitating conditions, perceived value, and ownership which play a crucial role in integrating gamified elements in FinTech apps [2]. Overall, gamification has the potential to bridge the gap between financial services and compliance and enhance the quality of the user experience.

**3-2- Social Influence Theory and The Technology Acceptance Model**

Two important theories are the underlying base for reconstructing business platforms with gamified elements. The first one social influence theory developed by Kelman who believed that an individual's attitude, beliefs, and behaviors are influenced by referent others through three processes and those are: compliance, identification, and internalization. According to this, social influence causes changes both in attitude and actions. This alteration occurs at different stages and levels [18]. Following the theory, "the satisfaction derived from compliance is due to the social effect of accepting the influence." Therefore, compliance is assumed to occur when individuals react positively and adopt the behavior of gaining rewards.

In the second stage of this process, identification occurs. This is when individuals project a behavior to maintain a desired coloration in their current circle. Hence, "the satisfaction occurs due to the act of conforming."

Lastly in this process, it is known that each individual adopts the proposed behavior knowing that it is congruent with their value system [18]. In this case, therefore, the satisfaction occurs due to "the content of the new behavior." This means that adaptation of such an act leads to a varied form of reward. Every single stage in this trio has a determinant of influence. Those are (a) the relative importance of the anticipated effect, (b) the relative power of the influencing agent, and (c) the prepotency of the induced response. It is important to note that, for each process, these determinants are qualitatively different. So, each process has a distinctive set of antecedent conditions; similarly, each process leads to a distinctive set of consequent conditions [15].

From the other side of the spectrum, the theory of technology acceptance model describes how the users accept and start to use a new technology. The main influential factors in this process are perceived usefulness (PU) and perceived ease-of-use (PEOU). These two perceptions are formed by the users during their initial encounters and if the perceptive outcome is positive the behavior is effectively adopted [19].

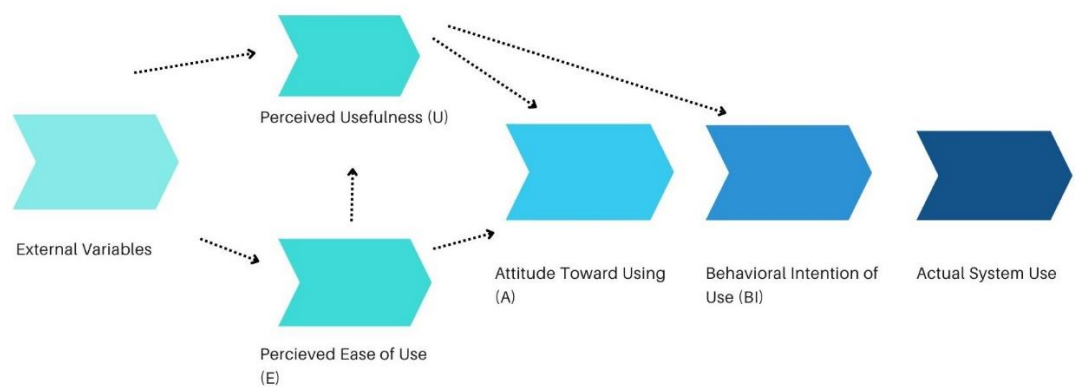

Figure 1. Technology Acceptance Model

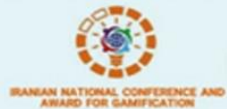

### 3-3-The Flow Theory

The correlation between ability and mastery has been a subject of research for several years, as highlighted in Caillois' work dating back to 1962 [12]. The Flow Theory postulates that the acquisition of ability and mastery is contingent upon learning. This occurs when there is an uninterrupted flow in the learning process and the focus is intensified.

According to the Theory of Flow, intense engagement in the activity is essential and is directly linked to the individual's proficiency level in the given context [12]. To attain a flow state where the user is completely absorbed in the activity, three fundamental conditions are necessary [12] : 1) a well-defined set of goals that guides the attention towards a single objective, 2) a balance between the difficulty of the task and the user's perceived ability to focus on the task, and 3) Clear and prompt feedback that guides the user on how to adjust their actions [13].

### 4. Adaptation of Gamification in the Fintech Platforms

The social influence theory, developed by Herbert Kelman, offers a framework for understanding how the environment around individuals can affect their behavior. The theory proposes three primary types of social influence: compliance, identification, and internalization [11]. In other words, the environment, including virtual ones, can significantly impact customer satisfaction and bridge the gap between financial services and compliance by shaping individuals' beliefs, attitudes, thoughts, and actions.

To improve the delivery system of Fintech applications, it is important to observe customer value from a service-related perspective. However, depending on the context, gamification components may need significant adaptation. Perceiving customer value, especially in the complex context of financial services, can be challenging and must be applied contextually within a specific framework [5]. One effective approach involves incorporating gamified elements with metaphorical verbs and deep-rooted cultural cues. For example, bestowing coupons with the presentation of red envelopes can improve the customer relationship with financial procedures experienced in software services and apps. Such examples can increase customer awareness of the brand and enhance their overall experience.

### 4-1- The Iterative stages of gamifying the Business Platform

The platform uses gamification to engage customers by offering a game-like experience. This experience is aimed at the active audience of the platform and has six objectives: absorption, enjoyment, activation, creative thinking, absence of negative affect, and dominance [8]. These objectives are supported by three theories: Flow Theory, Theory of Planned Behavior, and Technology Acceptance Model. Ensuring the effectiveness of the gamified experience is often overlooked. By correlating these objectives with the three theories, a better framework for developing a Fintech platform with effective gamified elements can be created [10].

Psychological ownership plays a vital role in designing gamified cycles. It encompasses self-efficacy, accountability, belongingness, and self-identity. Gamified components in the life cycle of a fintech app should correspond to this subset definition of the ownership factor [9].

There are four underlying motives behind a gamified designed element that corresponds to these four factors: efficacy and reflectance, self-identity, place, and stimulation. Carefully designed gamified elements that surrender to these motives can have a higher impact. The integration of gamified elements into the platform should be done subtly and at different stages of bilateral interaction between the user and the platform. The integration manner is important, and the user's attitude towards a brand, congruity, and prominence in the integration manner should be considered. Intrusiveness should be minimized as it weakens the initial relationship between the potential users and the app life cycle [9]. Low invasiveness leads to higher degrees of positive responses from users.

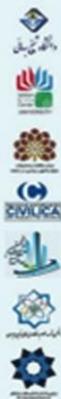
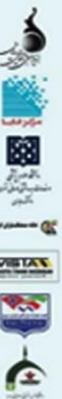

Two important data to increase the effectiveness of the gamified element are total fixture duration and visual attention. The gamified element should be in line with the narrative of each task to avoid intrusiveness [9]. A successful gamified element invites users to be continuously attentive to react accurately to the stimuli and finish the task at hand. The novelty of the experience is easily missed in the Fintech platform due to the nature of task narratives. Gamified n-back task cycles can re-stimulate the motivation seed in current users.

**4-2- The Gamified BP lifecycle**

In the design of the gamified components, mechanics describe the particularity and feature of this individual element. On the other hand, the dynamic indicates the run-time behavior of the mechanics in a gamified platform. Examples of such elements could be the status and progression over time [8].

Gamified elements have certain mechanics and those are the setup, rule, and progression mechanics. These are the posterior factors to consider in the design of gamified elements.[9] Another aspect of the gamified element in the non-game context is its aesthetics. in the frame of the gamified element, aesthetics correlate with the emotional responses evoked in the users while interacting with the fintech platform.

Action, social, mastery, achievement, immersion, and creativity aspect are the sixth pillar motives for playing a game. This is justified by the psychological studies of motivation and traditional designer theories [16]. This can be valid for the gamification of components in the non-game context.

Table 1. Aspects of Gamified BPM lifecycle

| An example of a column heading | Definition |
|---|---|
| Action Aspects | Responds to excitement |
| Social Aspects | Responds to community |
| Mastery Aspects | Responds to strategy learning and building |
| Achievement Aspect | Responds to the completion of the tasks |
| Immersion Aspects | Responds to the fluidity of the process |
| Creativity Aspects | Responds to the design of the elements |

On the higher definition, these motives can be clustered. The clustering is done based on the correlations. A type of clustering is as follows [16]:

Table 2. Motivational Clusters

| An example of a column heading | Encompasses | Theory Alignment |
|---|---|---|
| Action–Social Cluster | Both action and social aspects. | Social influence Theory |
| Mastery–Achievement Cluster | Both mastery and achievement aspects. | Flow Theory |
| Immersion–Creativity Cluster | Both immersion and creativity aspects. | Flow Theory |

**4-3- The Eight stages of the Gamified BP lifecycle**

The gamified business platform has its unique iterative cycle consisting of 8 episodes that need to be followed. This cycle is a result of the integration between the gamification and business platform life cycles. The novel cycle opens up opportunities for businesses to leverage gamification with greater accuracy and effectiveness [8].

The gamified business platform begins with an investigation of the technical environment and business processes. This initial analysis should identify any gaps or ambiguities. Based on this overview, a clearer understanding of what gamification should accomplish will be provided. The next phase involves identifying the participants. To maximize the impact of gamification in the business context, all potential users and their possible identification and active professional environments should be recognized.

The third episode in this cycle involves identifying the mission [8]. Each gamified element should have an appropriate mission. Identification of the objective contributes to a better understanding of the user's motivation and interest. The aim is to create a system based on a game mechanism that enhances the user's enthusiasm and interest.

The fourth step involves revealing the underlying motivation based on the theory of flow and self-determination. This contributes to developing an effective and comprehensive gamified strategy.

In the fifth stage, the game's mechanism should be identified based on the information gathered from the previous steps. The mechanism of the gamified element in the fintech platform could be the monthly financial interest ranking or the investment strategy badges. These are the visible aspects of the gamified platform. Once the mechanics have been defined, the dynamics and aesthetics will be generated to contribute to the gamified component's emotional invocation and run-time behavior.

The sixth step involves the design procedure of the gamified components [10]. To prepare a model in the graphical notation, both the objective and all the details from the previous stages should be taken into account. The design procedure should ensure that there are no undesirable aspects in the gamified component.

The implementation step requires verification of the design phase. The execution instances will be checked to evaluate the gamified Fintech platform.

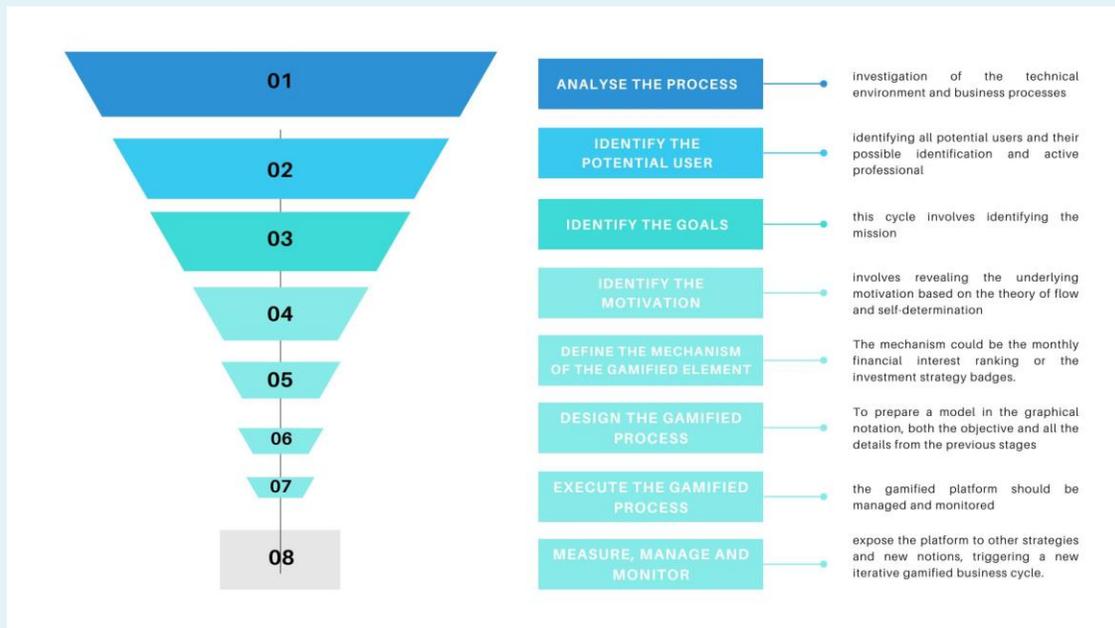

Figure 2. Gamified BPM lifecycle

The seventh step is not the final stage. After implementation, the gamified platform should be managed and monitored. Consequently, the eighth and last phase calls for evaluation and further control of the gamified Fintech platform. This step may expose the platform to other strategies and new notions, triggering a new iterative gamified business cycle.

**4-4- Cooperative Gamified Elements**

It is worth noting that participants continuously evaluate the potential outcomes of gamified structures. In the past, gamified elements were added systematically to a system. Exploring the design and implementation process of cooperative game theory, based on game theory models, can yield rich insights [13]. Effective task performance involves a process of experience, reflection, conceptualization, and testing. When interacting with a gamified item, this process should be integrated into the learning procedure [4]. Double-loop gamified elements focus on the conceptualization of a step-in terms of both process and cognition. However, learning occurs more chaotically, so the dynamic and flexible aspects of the involved gamified components matter greatly. When designing an element using game theory, it is crucial to focus on a specific aspect of the problem, whether it is procedural or solely dependent on the outcome. Another major dimension to the design of gamified elements is determining whether a particular element contributes to cooperativeness or competitiveness. Cooperative design encourages individuals to achieve a common task, while competitive design involves competing with one another for a single objective. Both approaches have their benefits, and extrinsic rewards can encourage competitive cooperation [10].

The integration of gamified components should be sensitive to insights about decision-making leading to concrete action [22]. Beforehand, the integration of gamified components can be visually represented with color-coded procedural trees, which better manifest the correlation level of the involved components. The intended use case of the game theory model provides a more accurate structure and leads to a rule-based integration of all the involved elements[10].

**4-5- Assistantship of the Audio-Gamified Components**

Audio gamification has great potential for enhancing user engagement and user experience. This can be achieved by integrating voice assistance in various platforms. Audio gamification is particularly effective in delivering gamified elements that do not require high processing units or specific visual representations. Sound effects can have an equal, if not greater, impact on users. However, implementing the right level of functionality and aesthetics poses certain challenges [6].

Deci and Ryan's self-determination theory (SDT) is an important theory in the design of gamified audio elements[6]. According to this theory, the main focus of the design of a gamified audio element should be on extrinsic motivation. This type of motivation leads to measurable and definable results and stimulates the user's motivational acts. For instance, audio assistance in Fintech platforms can create emotions and transmit information. To achieve this, assisted audio feedback in a task-oriented procedure can be provided. The design of audio-gamified elements requires careful consideration of auditory textures, especially in the context of simple and short messages [17]. Responses received through back-and-forth interactions with gamified audio elements can indicate the user's mood and sentiment. Adapting the context of a given platform requires a narrative that identifies the significant steps in a procedure.
Saving time is another way to leverage audio assistance and keep users motivated. This can encourage users to obtain more information in the context of personal portfolio investment development [17].

## 5. Evaluation Metrics for the Designed Elements

Variables, hypotheses, and planned analyses are all part of a successful design of a gamified element. Users have an initial perception of the game's designed elements. This association can be measured with relevant metrics such as effort degree [1]. Two other metrics for evaluating the gamified components are Cronbach's alpha coefficient and Composite reliability.

Cronbach's alpha coefficient is a metric used for measuring internal consistency and to scale the reliability of items based on design decisions. In this presented formula $P_T$ stands for tau-equivalent reliability, $k$ stands for the number of items, $\sigma_{ij}$ stands for covariance between $X_i$ and $X_j$ and $\sigma_X^2$ stands for item variance and inter-item covariances.

$$P_T = \frac{k^2 \overline{\sigma_{ij}}}{\sigma_X^2}$$

Composite reliability refers to measuring the internal consistency of overall items. $P$ stands for the number of indicators, $\lambda_i$ is the standardized loading for the *ith* indicator and $V(\sigma)$ variance of the error term for the *ith* indicator.

$$\frac{\left(\sum_{i=1}^{p} \lambda_i\right)^2}{\left(\sum_{i=1}^{p} \lambda_i\right)^2 + \sum_{i}^{p} V(\sigma)}$$

The range and the optimum value for these metrics are as follows:

Table 3. Measurement Metrics for Gamified Elements

| Measurement Metrics | Range | Optimum Results Range |
|---|---|---|
| Cronbach's alpha | 0 to 1 | $0.80 > \alpha > 0.70$ |
| Composite reliability | 0 to 1 | $0.70 < \alpha$ |
| Average variance extracted | Interchangeable | $0.50 < \alpha$ |
| Average covariance extracted | Interchangeable | Large positive |

Facilitating the effort expectancy and enhancing the perceived value leads to higher intention and a gamified procedure or carefully designed Fintech platform with gamified elements can accord with this vision [22]. These evaluation measurements can be used once the items are in the first round of integration use.

## 6. Conclusion

Gamified elements in financial services provide both opportunities and challenges. Following the proposed notions, we can implement a novel life cycle surrounding two life cycles of gamified Fintech portfolio management and Fintech procedure development. This leads to Increasing the diversity of the investment portfolio need. The combination of the motivational factors can proceed to actualizing and finalizing the financial act. Moreover, the eight aspects of gamified element construction have been presented in accordance with the theories supporting this expansion. Integrating gamified elements into business platform such as Fintech opens more doors for individuals to invest effectively and take part in wider range of investment opportunities.

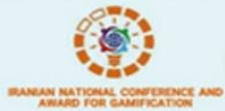

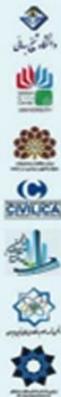
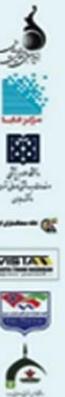